\documentclass[sigconf]{acmart}

\usepackage{booktabs} 
\usepackage{amssymb,amsmath}
\usepackage{algorithm}  
\usepackage{algpseudocode}   
\usepackage{caption}
\usepackage{subfigure}
\usepackage{textcomp}

\usepackage{balance}

\def\BibTeX{{\rm B\kern-.05em{\sc i\kern-.025em b}\kern-.08emT\kern-.1667em\lower.7ex\hbox{E}\kern-.125emX}}
    
\begin{document}

\fancyhead{}

\copyrightyear{2019} 
\acmYear{2019} 
\setcopyright{acmcopyright}
\acmConference[KDD '19]{The 25th ACM SIGKDD Conference on Knowledge Discovery and Data Mining}{August 4--8, 2019}{Anchorage, AK, USA}
\acmBooktitle{The 25th ACM SIGKDD Conference on Knowledge Discovery and Data Mining (KDD '19), August 4--8, 2019, Anchorage, AK, USA}
\acmPrice{15.00}
\acmDOI{10.1145/3292500.3330686}
\acmISBN{978-1-4503-6201-6/19/08}

\title{IntentGC: a Scalable Graph Convolution Framework Fusing Heterogeneous Information for Recommendation}


\author{Jun Zhao}
\email{jun.zhaozj@alibaba-inc.com}
\author{Zhou Zhou}
\email{xiaoyuan.zz@alibaba-inc.com}
\affiliation{%
  \institution{Alibaba Group}
}

\author{Ziyu Guan}
\authornote{Corresponding author.}
\email{zyguan@xidian.edu.cn}
\affiliation{%
  \institution{Xidian University}
}

\author{Wei Zhao}
\email{ywzhao@mail.xidian.edu.cn}
\affiliation{%
  \institution{Xidian University}
}

\author{Wei Ning}
\email{wei.ningw@alibaba-inc.com}
\affiliation{%
  \institution{Alibaba Group}
}

\author{Guang Qiu}
\email{guang.qiug@alibaba-inc.com}
\affiliation{%
  \institution{Alibaba Group}
}

\author{Xiaofei He}
\email{xiaofei.he@cad.zju.edu.cn}
\affiliation{%
  \institution{Zhejiang University}
}

%
\renewcommand{\shortauthors}{J. Zhao et al.}

%
\begin{abstract}
The remarkable progress of network embedding has led to state-of-the-art algorithms in recommendation. However, the sparsity of user-item interactions (i.e., explicit preferences) on websites remains a big challenge for predicting users' behaviors. Although research efforts have been made in utilizing some auxiliary information (e.g., social relations between users) to solve the problem, the existing rich heterogeneous auxiliary relationships are still not fully exploited. Moreover, previous works relied on linearly combined regularizers and suffered parameter tuning.

In this work, we collect abundant relationships from common user behaviors and item information, and propose a novel framework named IntentGC to leverage both explicit preferences and heterogeneous relationships by graph convolutional networks. In addition to the capability of modeling heterogeneity, IntentGC can learn the importance of different relationships automatically by the neural model in a nonlinear sense. To apply IntentGC to web-scale applications, we design a faster graph convolutional model named IntentNet by avoiding unnecessary feature interactions. Empirical experiments on two large-scale real-world datasets and online A/B tests in Alibaba demonstrate the superiority of our method over state-of-the-art algorithms.
\end{abstract}

%
%



%
\keywords{Graph Convolutional Networks, Recommendation, Heterogeneous Information Network}

%

%
\maketitle

\section{Introduction}

With the ever-growing volume of online information, recommender system has become an effective key solution on a variety of websites (e.g., Amazon, Youtube, Alibaba) for helping users discover interesting products or contents. Due to successes of deep learning and network embedding in recent years, the algorithms that power these recommender systems are generally based on the idea that user preferences or item semantics can be learned by neural models in the form of low-dimensional representations, which in turn can be used for recommendation by searching the closest embeddings in the low-dimensional space. 

Among different information that could be obtained on websites, user interactions with items (clicks, etc.) are the most common and explicit indicators of user preferences. Many algorithms have been proposed by utilizing these explicit behaviors to predict users' preferred items \cite{xu2016tag,he2017neural}. However, a major downside is that, these explicit preferences are quite sparse, which severely limits the model capability for recommendation. On the other hand, there are usually rich auxiliary relationships that imply user preferences and item semantics, which could help overcome the sparsity issue. Several research works have explored such auxiliary relationships and demonstrated their effectiveness \cite{elkahky2015multi,wang2018billion,gao2018bine}. To list a few, Wang \textit{et al.} \cite{elkahky2015multi} proposed a cross-domain solution that preserve user-user social relationships from a social domain and user-item relationships from a content domain. The authors in \cite{wang2018billion} modeled items in a homogeneous graph and adopted a DeepWalk approach to preserve item co-occurrences (clicked in the same session). These auxiliary relationships widely exist and can be used to improve the performance of recommendation.

\begin{figure} 
\includegraphics[height=1.4in, width=2.7in]{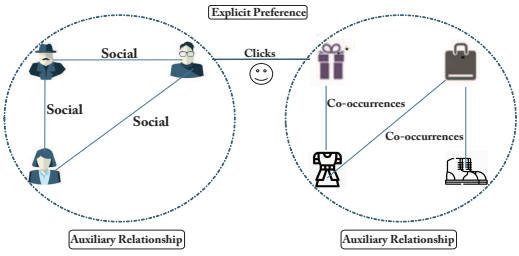}
\caption{Examples of previously utilized auxiliary relationships}
\label{previous}
\end{figure}
\begin{figure} 
\includegraphics[height=1.8in, width=3.4in]{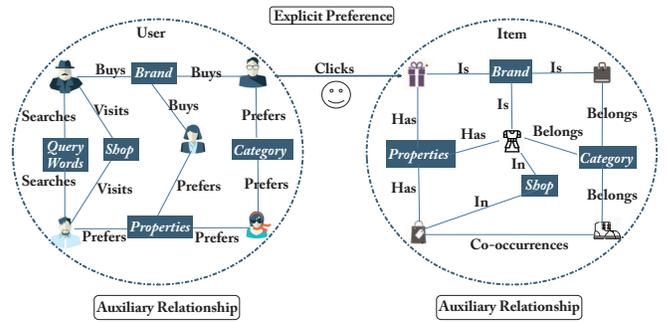}
\caption{Examples of heterogeneous auxiliary relationships on e-commerce websites}
\label{intro}
\end{figure}

However, we find that all previous works only captured one type of auxiliary information for users and/or one type for items in the model (see Fig.~\ref{previous}), while ignoring a plenty of additional heterogeneous relationships on the graph. We provide an illustrative example on an e-commerce website in Fig.~\ref{intro}. We can see that, besides explicit user interactions on items, there is rich auxiliary information such as user submitted query words, visited shops, preferred brands and properties. These auxiliary relationships are potentially useful in capturing more semantics and relevance. For instance, the query words contain content information of user requirements which are effective to link users with similar interests as well as to find items with similar content. Likewise, the brands link users with similar taste of fashion styles and provide complementary information to content similarity. However, these heterogeneous auxiliary relationships are not fully considered in recommendation. In this work, we are concerned with studying a unified framework to capture both explicit preferences and all heterogeneous auxiliary relationships of users and items.

To this end, we extend Graph Convolutional Network (GCN) to achieve the goal. The core idea behind GCN is to generalize the convolutional neural networks on graph-structured data \cite{kipf2016semi,hamilton2017inductive}, which has presented capabilities of content propagation and high-level expressiveness, and demonstrated great success in node classification tasks. More recently, researchers in Pinterest have adopted GraphSage (a state-of-the-art GCN model) on an item graph to recommend related items \cite{ying2018graph}. However, both their problem and solution are intrinsically different from our work due to:

1) Their model considers only item information, while ignoring users and auxiliary objects.

2) To scale up, GraphSage needs to sample many clustered mini-graphs of items for embedding reuse. However, it is hard to find such clustered mini-graphs that contain both users and items, due to the sparsity issue mentioned above. It is very likely that the mini-graph sampling algorithm ends up with a very large subgraph (or even the whole graph). Thus, the idea of GraphSage is not suitable for large-scale user-item graphs in our context.

3) Their method is proposed for homogeneous networks, while user-item graphs studied in this work are heterogeneous.

\vspace{1mm} \noindent \textbf{Our Work}. In this work, we propose a novel GCN-based framework called IntentGC for large-scale recommendation which captures both explicit user preferences and heterogeneous relationships of auxiliary information by graph convolutions. There are mainly three innovative points of IntentGC: 

1) Fully exploiting auxiliary information: We capture plenty of heterogeneous relationships to boost the performance of recommendation. To facilitate modeling and improve robustness, we translate auxiliary relationships of first order proximity into more robust weighted relationships of second order proximity. For instance, if user1 submits a query word ``Spiderman'', we think there is a connection between user1 and ``Spiderman'' (first order proximity). If user1 and user2 both submit query words ``Spiderman'', ``Ironman'', ``Thor'', we think there is a more robust relationship between user1 and user2 (second order proximity), as they are probably fans of Marvel movies. With different types of auxiliary objects, we can generate heterogeneous relationships of second-order proximity. IntentGC automatically determines the weights of different types of relationships in training. We find these heterogeneous relationships are useful and complementary to each other in practice, and can significantly improve the performance.

2) Faster graph convolution: To remove the limitation of training on clustered mini-graphs for large-scale graphs, we propose a novel convolutional network named IntentNet, which is not only more efficient but also more effective than GraphSage. The IntentNet takes a faster graph convolution mechanism. The key idea of IntentNet is to avoid unnecessary feature interactions by dividing the functionality of graph convolution into two components: a vector-wise convolution component for neighborhood feature propagation and a fully-connected network component for node feature interaction. 
Benefiting from this mechanism, we deploy a distributed architecture for simple mini-batch training (sampling nodes).

3) Dual graph convolution in heterogeneous networks: To preserve the heterogeneity of users and items, we design a dual graph convolutional model for network representation learning. First, we take advantage of two independent IntentNets that separately operate on user nodes and item nodes. After nonlinear projection through the fully-connected network in the respective IntentNet, the obtained embeddings of users and items can be deemed to form a common space. Then, with training guided by explicit preferences, relevance can be assessed between users and items in the space.


It is worth to note that, unlike previous works of capturing auxiliary relationships in the objective function with a regularizer \cite{elkahky2015multi,gao2018bine}, which is linear and heavily depends on handcraft parameter tuning, our method can automatically learn the importance of different auxiliary relationships through non-linear neural network. We note that auxiliary information could also be designed as node input features. However, nodes sharing some input features would not be near in the high-level embedding space due to the complex neural network projection. By further modeling auxiliary information as relationships in translated affinity graphs, IntentGC can directly learn from node relationships, which could significantly improve the performance. Experiments also confirm it.

The main contribution of this work is summarized in the following: 

1) We propose IntentGC, an effective and efficient graph convolution framework. To our best knowledge, this is the first work to model explicit preferences and heterogeneous relationships in a unified framework for recommendation.


2) We design a novel graph convolutional network named IntentNet with a faster graph convolution mechanism. It leads to 22.1\% gain in MRR and 75.6\% reduction of running time. 

3) We conduct extensive offline experiments on two large-scale datasets and deploy an online system with production A/B tests at Alibaba. In offline evaluation, we improve MRR by 95.1\%, and in online A/B tests IntentGC shows 65.4\% improvement in click-through-rate (CTR), compared to the best baseline.


\section{Related Work}

\subsection{Network Embedding}
Network embedding aims to represent graph nodes in a low dimensional space where the network structure and properties are preserved. It can benefit a variety of tasks including recommendation. Many effective network embedding algorithms have been proposed  \cite{perozzi2014deepwalk,grover2016node2vec,tang2015line,wang2016structural,zhu2018deep}. We briefly review some of these methods here. Readers can refer to \cite{cui2018survey,goyal2018graph} for a comprehensive survey. DeepWalk \cite{perozzi2014deepwalk} deployed truncated random walks on nodes to learn latent representations by treating walks as the equivalent of sentences. Following this pioneer work, node2vec \cite{grover2016node2vec} extended DeepWalk with more sophisticated random walks and breadth-first search schema. SDNE \cite{wang2016structural} exploited the first-order proximity and second-order proximity in a joint approach to capture both the local and global network structure. DVNE \cite{zhu2018deep} learned a Gaussian distribution in the Wasserstein space for each node to preserve more properties such as transitivity and uncertainty.


While extensive works have been made for representation learning in homogeneous networks, the graphs in real-world applications are more likely to be heterogeneous information networks (HINs). To take advantage of the rich information in HINs, a couple of algorithms have also been proposed to deal with the heterogeneity \cite{dong2017metapath2vec,shi2018easing,chang2015heterogeneous,zhang2018deep}. Metapath2vec++ exploited meta-path based random walks to maximize the biased transition probabilities according to human expertise \cite{dong2017metapath2vec}. HEER embeded HINs via an additional edge representation to bridge the semantic gap between heterogeneous nodes \cite{shi2018easing}. Although these methods can be applied to general HINs, they treat each type of relationships with equal importance in the model, which is not appropriate to boost recommender systems since the user-item relationships are the main objective to predict. Currently, little attention has been paid to utilizing heterogeneous information to strengthen the performance of recommendation.


\subsection{Graph Convolutional Networks}
In recent years, more attention was paid to applying convolutional neural networks on graph-structured data \cite{bruna2013spectral,kipf2016semi,hamilton2017inductive,velickovic2017graph}. Bruna \textit{et al.} defined the convolution operation in the Fourier domain by computing the eigendecomposition of the graph Laplacian \cite{bruna2013spectral}. To reduce the complexity of convolution, Kipf and Welling proposed the GCN model by simplifying previous methods via first-order approximation of localized spectral filters on graphs \cite{kipf2016semi}. To be specific, they considered a convolution operation for each node as a mean-aggregation of all the adjacent feature vectors, with a transformation by a fully connected layer and a nonlinear activation function. However, in their model, the representations of the central node and neighbor nodes were aggregated with the same weight and non-trainable. More recently, Hamilton \textit{et al.} proposed GraphSage that extends the GCN approach further in an inductive manner \cite{hamilton2017inductive}. This technique sampled a fixed-size neighborhood of each node to avoid operating on the entire graph Laplacian. They also improved GCN by concatenating each node's representation with the aggregated neighborhood vector for learning interaction weights. Although there is remarkable progress, previous research works in GCN are mainly focused on homogeneous graphs. In this work, we propose a novel algorithm that extends GCN to heterogeneous information networks, and significantly improves both effectiveness and efficiency in recommendation.


\subsection{Recommendation}
Recently, deep learning based algorithms have achieved significant successes in the recommendation literature \cite{zhang2017deep}. Based on whether capturing user information in the model, there are mainly two types of methods: 1) item-item recommendation and 2) user-item recommendation. The motivation of item-item recommendation is to find similar items to a user's historically interacted items. In this line of works, Wang and Huang \textit{et al.} \cite{wang2018billion} adopted a DeepWalk approach with side information to obtain vector representations on an item graph. Ying \textit{et al.} \cite{ying2018graph} proposed a random walk based GraphSage algorithm (named PinSage). 


Different from the above works, our method falls into the user-item recommendation group \cite{he2017neural,xu2016tag,lian2018xdeepfm}. This group of methods aims to predict a user's preferred items directly, which is generally more related to satisfaction of users and also more challenging due to the sparsity issue. To alleviate the problem of sparsity, several works attempt to utilize additional auxiliary relationships. For example, Gao \textit{et al.} \cite{gao2018bine} designed a biased random walk method to utilize deduced user-user and item-item relationships from the user-item bipartite graph. Wang \textit{et al.} \cite{elkahky2015multi} incorporated social relationships in a cross-domain setting. However, existing methods consider only one type of auxiliary relationships for users and/or items, while ignoring abundant heterogeneous auxiliary relationships on the graph. Moreover, previous methods usually capture auxiliary relationships via regularizers, which limits the capability of the model and also heavily depends on handcraft parameter tuning. In this work, we propose a novel framework IntentGC to exploit both explicit preferences and a rich set of heterogeneous auxiliary relationships. It can automatically determine the importance of different kinds of auxiliary relationships by graph convolutions.

\section{Problem Definition}  
\label{sec:defn}
We mathematically formulate the problem of recommendation. First, let us consider a typical scenario on an e-commerce website: Last week, Jack has queried about a couple of keywords with some demands. From the lists of returned items, he clicked some attractive items for detailed information. During this week, he also visited some online shops for checking out new books. Finally on Saturday, he purchased several books with the bestseller property and a T-shirt of his favorite brand. Based on Jack's behaviors, the platform has collected rich information (submitted query words, clicked items, visited shops, preferred properties and brands) for recommending potential interesting items to him in a personalized manner.

This kind of recommendation scenario could also be observed on other websites. Generally, multiple kinds of objects and historical user behaviors on a website form a heterogeneous information network, as defined in the following:

\newtheorem{hin}{Definition}
\begin{hin} \label{hin: defn}
\textbf{Heterogeneous Information Network} (HIN) is an undirected graph $\mathcal{G} = (\mathcal{V}, \mathcal{E})$, where $\mathcal{V}$ is the set of nodes and $\mathcal{E} \subseteq \mathcal{V} \times \mathcal{V}$ is the set of edges between nodes in $\mathcal{V}$. $\mathcal{G}$ is associated with a node type mapping $\varphi: \mathcal{V} \rightarrow \Gamma^{v}$ and an edge type mapping $\psi: \mathcal{E} \rightarrow \Gamma^{e}$, where $\lvert \Gamma^{v} \rvert > 1$ and / or $\lvert \Gamma^{e} \rvert > 1$. $\mathcal{V}$ can be written as $\mathcal{V} = \mathcal{V}_1 \cup \mathcal{V}_2 \cup \dots \cup \mathcal{V}_r \cup \dots \cup \mathcal{V}_R $, where $\mathcal{V}_r$ denotes the set of nodes with type $r$ and $R=\lvert \Gamma^{v} \rvert$.  
\end{hin}

\noindent \textbf{User-Item Recommendation}. Particularly, in recommendation we denote $\mathcal{V}_1$ as the set of user nodes and $\mathcal{V}_2$ as the set of item nodes, with $\mathcal{V}_3, \dots ,\mathcal{V}_R$ representing the other objects' nodes (query words, brands, etc.). We also denote $\mathcal{E} = \mathcal{E}_{label}\cup \mathcal{E}_{unlabel} $, where $\mathcal{E}_{label} \subseteq \mathcal{V}_1 \times \mathcal{V}_2$ represents the set of edges between user nodes and item nodes, and $\mathcal{E}_{unlabel} = \mathcal{E} \setminus \mathcal{E}_{label}$ represents the other edges. Since a typical recommendation setting in real-world is to predict a user's preferred items according to previous behaviors, we use $\mathcal{G} = (\mathcal{V}, \mathcal{E})$ to denote the graph constructed by historical data, and $\mathcal{G}^p = (\mathcal{V}^p, \mathcal{E}^p)$ to denote the graph of the real future. Then we can formulate the user-item recommendation problem as a link prediction problem on graph in the following:

\textbf{Input}: A HIN $\mathcal{G}= (\mathcal{V},\mathcal{E})$ based on historical data.

\textbf{Output}: A predicted edge set $\widehat{\mathcal{E}}^p_{label}$, which is the prediction of the real edge set $\mathcal{E}^p_{label}$ on $\mathcal{G}^p$.

\section{Methodology}
In this section, we propose a novel framework for user-item recommendation on HIN. Our approach has three key characteristics: i) Network Translation, which translates the original graph into a special kind of HIN; ii) Faster Convolutional Network, which takes the advantage of vector-wise convolution for scaling up and synthesizes heterogeneous relationships in an optimal sense; iii) Dual Graph Convolution, which learns representations for both users and items on the translated HIN. Finally, we summarize the framework of our solution.

\subsection{Network Translation}
\label{sec:translation}
The heterogeneous nodes and relationships as the ones in Fig.~\ref{intro} provide us with not only rich information but also incompatible semantics and more challenges. Although modeling each type of edges using a type-specific manifold is a possible solution \cite{shi2018easing}, the high complexity and computational cost is infeasible for large data when dealing with many types of nodes and edges. Fortunately, in recommendation we only care about the representations of users and items. Motivated by this, we adapt a method similar to \cite{gao2018bine,zhang2018deep} to translate original auxiliary relationships to user-user relationships or item-item relationships. Intuitively, if users $u_1$ and $u_2$ are both connected by an auxiliary node in $\mathcal{V}_r (r>2)$, there is also an indirect relationship between $u_1$ and $u_2$. In this paper, we utilize the second-order proximity \cite{goyal2018graph} to capture the similarity between two users (or items), which is measured by the number of common auxiliary neighbors of the same type shared by them. In this way, we can encode the semantic information brought by auxiliary nodes into a set of heterogeneous user-user relationships and/or item-item relationships, and translate the HIN accordingly. Other generation method like meta-path based random walk is also applicable for network translation, but our approach leads to robust neighborhoods and simple implementation.




For clarity and ease of derivation, we first consider the case $\mathcal{V} = \mathcal{V}_1 \cup \mathcal{V}_2 \cup \mathcal{V}_3$. I.e., we only have one type of auxiliary nodes. By adding new second-order relationships to and removing the original auxiliary relationships and nodes from the HIN $\mathcal{G}$, we obtained a new and simplified HIN $G = (U, V, E_U, E_V, E_{label})$, where $U$ and $V$ denote the sets of user nodes and item nodes, respectively. $E_{label}  \subseteq U \times V$ is exactly the same as $\mathcal{E}_{label}$ before translation. $E_U  \subseteq U \times U$  and $E_V  \subseteq V \times V$ are the sets of generated edges between users and between items, respectively. Note for clarity, we assume there is only one type of edges between $\mathcal{V}_1$ (or $\mathcal{V}_2$) and $\mathcal{V}_3$. Nevertheless, our framework is general and allows multiple edge types. Each $e_{u}(i, j) \in E_U$ is associated with a similarity weight $s_{u}(i, j)$, which represents the second-order proximity in the origin graph. $s_{v}(i, j)$ for $e_{v}(i, j)$ is defined with the same notion as $s_{u}(i, j)$. Therefore, we can use $\mathbf{S}_{U} = [s_{u}(i, j)]$ and $\mathbf{S}_{V} = [s_{v}(i, j)]$ to represent the weight matrices for $E_U$ and $E_V$, respectively. We also define ${\mathcal{N}(u_i)}$ (or ${\mathcal{N}(v_i)}$), the set of neighbors of each node $u_i$ (or $v_i$), as the top $\rho$ within-type similar nodes according to $\mathbf{S}_U$ (or $\mathbf{S}_V$). 

Now we consider the case $\mathcal{V} = \lbrace \mathcal{V}_1, \mathcal{V}_2, \dots, \mathcal{V}_R \rbrace$ with $R$ types of nodes. For each auxiliary node type, we follow the process above for generating user-user/item-item edges. In this way, we can finally get $2R-4$ types of heterogeneous relationships where each can be denoted as $E_{U}^{(r)}$ (or $E_{V}^{(r)}$) with a weight matrix $\mathbf{S}_{U}^{(r)}$ (or $\mathbf{S}_{V}^{(r)}$). Likewise, the corresponding neighborhoods for $u$ and $v$ can be denoted by ${\mathcal{N}^{(r)}(u)}$ and ${\mathcal{N}^{(r)}(v)}$, respectively. 




We call the translated graph $G$ as user-item HIN. Then the problem becomes predicting $E_{label}^p$ (i.e., $\mathcal{E}_{label}^p$) given the user-item HIN $G$.

\subsection{Faster Convolutional Network: IntentNet}
\label{sec:intentnet}

\noindent \textbf{Motivation}. The core idea of GCNs is iteratively aggregating feature information from neighborhood by local filters. However, a major downside of it is the high complexity of computation. For instance, a 3-stacked GCN model with only 10 truncated neighbors involves 100+ convolution operations for each node on the graph, which is unacceptable for web-scale applications as the whole graph usually has hundreds of millions of nodes. In previous works, a common method for scaling up is to use a mini-subgraph sampling strategy, as did in \cite{ying2018graph}. They develop a producer-consumer distributed training method. In each iteration, it samples a clustered subgraph $\mathcal{M}$ by producer, and performs forward propagation on $\mathcal{M}$ to get all the nodes' representations by consumer. A clustered subgraph is produced in a breadth-first-search style on item-item graph. In this way, GCN is performed only once for each sampled subgraph with embedding vectors being reused during updating (all training pairs should be contained in the subgraph). However, for user-item HIN $G$, it is difficult to generate such clustered subgraphs for representation reusing. This is because the user-item preference links are quite sparse. If we follow the producer in their method for sampling, we would get a very huge subgraph, or even the whole graph. Hence, in order to apply our approach on large-scale graphs, we develop a faster convolution operation which allows ordinary node sampling.


\vspace{3mm} \noindent \textbf{Vector-wise convolution operation}. For clarity, we first consider only one type of auxiliary relationships, and then extend our method to handle heterogeneous relationships. We only use user nodes for illustration since user nodes and items nodes are symmetric on $G$. A layer of graph convolution contains two parts: 1) aggregation and 2) convolution function. The aggregation is a pooling layer for aggregating the feature information from neighbors, which can be formulated as:
\begin{equation} \label{eq:aggregation}
\mathbf{h}_{\mathcal{N}(u)}^{k-1} = \textrm{AGGREGATE}({\mathbf{h}_a^{k-1},  \forall a \in \mathcal{N}(u)})
\end{equation}
where $\mathbf{h}_a^{k-1}$ denotes the embedding vector of user $a$ after the $(k-1)$-th convolutional layer, and $\textrm{AGGREGATE}$ is a pooling function (mean, etc.). The neighborhood vector $\mathbf{h}_{\mathcal{N}(u)}^{k-1}$ incorporates feature information from $u$'s neighborhood into the representation.

After aggregation, we need to combine the utility of self node and neighborhood by a convolution function. A typical convolution function is \cite{hamilton2017inductive}:
\begin{equation} \label{eq:concate}
\mathbf{h}_{u}^{k} = \sigma (\mathbf{W}^k  \cdot \textrm{CONCAT}(\mathbf{h}_{u}^{k-1}, \mathbf{h}_{\mathcal{N}(u)}^{k-1}))
\end{equation}
This function first concatenates the current node's vector $\mathbf{h}_{u}^{k-1}$ and its neighborhood vector $\mathbf{h}_{\mathcal{N}(u)}^{k-1}$. Then, it feeds the concatenated vector through a fully-connected neural network layer with nonlinear activation $\sigma$, so as to learn feature interactions during the representation transformation. We call this conventional function ``bit-wise'' in this paper.

\begin{center}
\begin{figure}
\subfigure[Bit-wise]
{ 
\begin{minipage}[t]{0.45\linewidth}
\includegraphics[height=35mm, width=35mm]{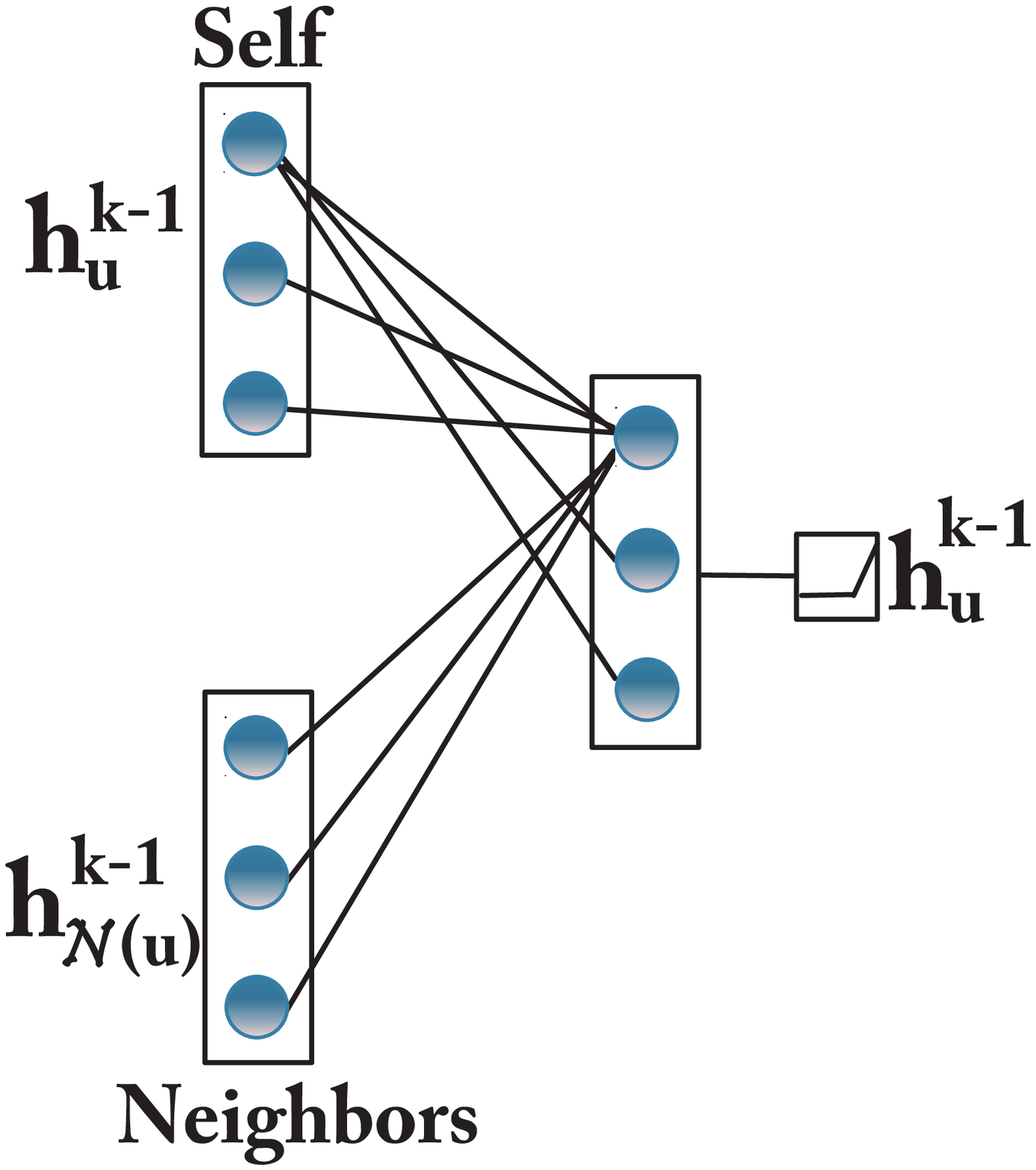}
\end{minipage}
}
\subfigure[Vector-wise]
{ 
\begin{minipage}[t]{0.45 \linewidth}
\includegraphics[height=35mm, width=45mm]{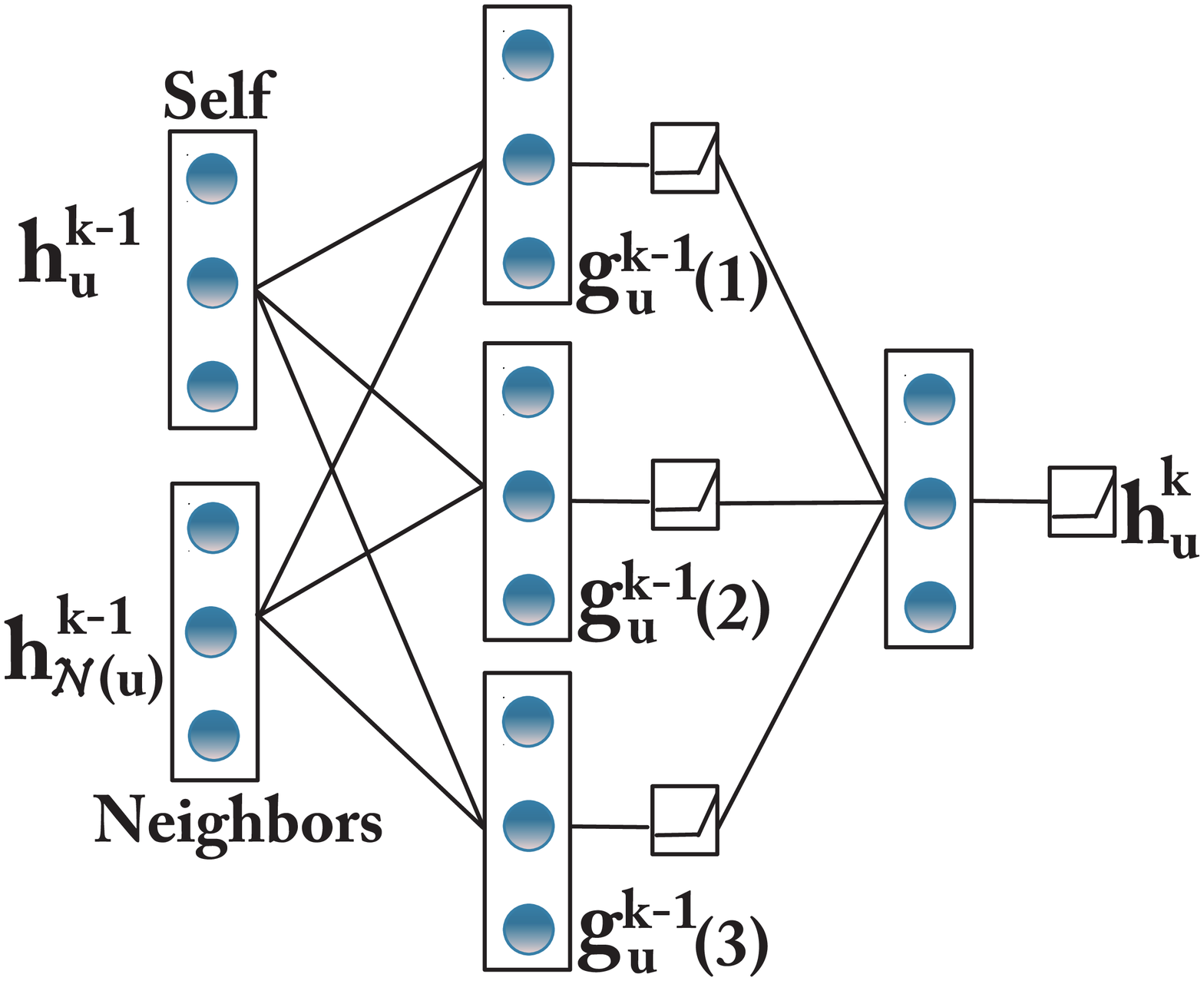}
\end{minipage}
}
\caption{Bit-wise and vector-wise graph convolution}
\label{fig:vectorwise}
\end{figure}
\end{center}

However, we observe that it is not necessary to learn all feature interactions between every pair of features in the concatenated vectors. During representation learning, there are mainly two tasks in the convolution operation: One is to learn the interactions between self node and its neighborhood, which determines how neighborhood boosts the results; the other one is to learn the interactions between different dimensions of the embedding space, which will extract useful combinatory features automatically. A key insight is that the interaction between feature $h_i$ in $\mathbf{h}_{u}^{k-1}$ and $h_j$ ($j \neq i$) in $\mathbf{h}_{\mathcal{N}(u)}^{k-1}$ is less informative. For instance, the age and career of a user (feature interaction in the same node) might suggest some preferred categories. Incorporating the rating feature of a user's neighborhood into his representation (interaction between nodes w.r.t the same feature) might be helpful to recommend related items. However, combining the age of a user and the rating feature of his neighborhood would probably result in meaningless guesses. Based on this observation, we designed a vector-wise convolution function in the following:
\begin{equation} \label{eq:vectorwise}
\mathbf{g}_{u}^{k-1}(i) = \sigma (w_u^{k-1}(i, 1) \cdot \mathbf{h}_{u}^{k-1} + w_u^{k-1}(i, 2) \cdot \mathbf{h}_{\mathcal{N}(u)}^{k-1})
\end{equation}
\begin{equation} \label{eq:kernel}
\mathbf{h}_{u}^{k} = \sigma (\sum_{i=1}^{L} \theta_{i}^{k-1} \cdot \mathbf{g}_{u}^{k-1}(i))
\end{equation}
where $w_u^{k-1}(i, 1)$ and $w_u^{k-1}(i, 2)$ denote the $i$-th local filter's weights for self node and neighborhood, respectively. Each local filter in Eq.~(\ref{eq:vectorwise}) can be viewed as learning how self node and neighborhood interact in a vector-wise manner. With all local filters being learned, we use another vector-wise layer as in Eq.~(\ref{eq:kernel}) to encode them into the representation of $\mathbf{h}_{u}^{k}$ for the next convolutional layer. The multiple local filters here ensure a rich information extraction capability, following the spirit of CNN \cite{Goodfellow-et-al-2016}. All these weights are shared on the graph. A comparison of vector-wise convolution and bit-wise convolution is depicted in Fig.~\ref{fig:vectorwise}. For either bit-wise or vector-wise approach, the graph convolution can be viewed as operations on $\mathbf{X}_k \in \mathbb{R}^{N \times M \times C}$, where $N$ denotes the number of nodes, $M$ denotes the number of neighbors, and $C$ denotes the dimensionality of node representation. If we view $C$ as the number of channels of the tensor $\mathbf{X}_k$, by basic mathematical derivation, bit-wise convolution is equivalent to 1-D CNN and our vector-wise convolution is equivalent to a variant of 1-D CNN where local filters' weights are shared among channels. We omit the detailed proof due to space limitation.

\noindent \textbf{IntentNet}. With the proposed convolution operation, we then can build stacked convolutional layers to form a network, which is highly efficient and capable of learning useful interactions from neighborhood propagation. However, this only achieves one task of graph convolution, so we further feed the output representation of the last convolutional layer through three additional fully-connected layers, in order to learn the feature interactions among different dimensions of the embedding space. We name this method as IntentNet, with the core idea of dividing the work of graph convolution into two components: vector-wise convolution for learning the neighborhood's utility, and fully-connected layers for extracting the node-level combinatory features. In practice, IntentGC is not only more efficient than conventional GCNs but also more effective in performance. A probable reason is that IntentGC can avoid useless feature interactions and is more robust to overfitting. More details will be presented in Section~\ref{sec:offline}.



\vspace{3mm} \noindent \textbf{Complexity}. Generally, we use $m$ to denote the sizes of representation vectors in different layers since they are similar in order of magnitude. Firstly, we analyze the complexity of the convolution operation. For $\rho$-neighborhood, each convolution operation of IntentNet needs an aggregation with $\mathcal{O}(m \ast \rho)$ complexity, a vector-wise scanning with $\mathcal{O}(m \ast L)$ complexity (L is the number of local filters) and a local filters merge with $\mathcal{O}(m \ast L)$ complexity. Putting them together, the time cost is $\mathcal{O}(m \ast (\rho + L))$. Since $\rho \ll m$ and $L \ll m$ are typically small integers\footnote{In our case, $\rho = 10$ and $L = 3$, while $m$ is at least several hundred, in order to well consume large-scale training data}, they can be regarded as constants. Then we have $\mathcal{O}(m \ast (\rho + L)) \approx \mathcal{O}(m)$. In comparison, GraphSage (using ordinary node sampling for training) needs an aggregation with $\mathcal{O}(m \ast \rho)$ complexity and a dense convolution (Eq.(\ref{eq:concate})) with $\mathcal{O}(m^2)$ complexity. The total cost is $\mathcal{O}(m \ast (\rho + m)) \approx \mathcal{O}(m^2)$ for each convolution operation. Now we analyze for $q$-stacked graph convolution. Both algorithms will run $q$ iterations, where the $r$-th iteration needs $(1+\rho+\cdots+\rho^{q-r}) = \frac{\rho^{q-r+1} - 1}{\rho-1}$ number of graph convolutions. For $q$ iterations, both algorithms need $\frac{\rho+\rho^2+\cdots+\rho^q-q}{\rho-1} \approx \rho^{q-1}$ number of graph convolutions. In summary, IntentNet's time cost is $\mathcal{O}(\rho^{q-1} \ast m + m^2)$ (the $m^2$ term corresponds to the cost of the fully-connected layers after the $q$ convolutional layers), while GraphSage's cost is $\mathcal{O}(\rho^{q-1} \ast m^2)$. We conclude that IntentNet is more efficient than GraphSage.





\vspace{3mm} \noindent \textbf{Heterogeneous relationships}. We now extend IntentNet to capture more heterogeneous relationships of auxiliary information in the model. Consider $E_U = E_U^{(1)} \cup E_U^{(2)} \cup \dots \cup E_U^{(R-2)} $ with $R-2$ types of user-user relationships. In this case, the vector-wise convolution operation in Eq.~(\ref{eq:vectorwise}) can be generalized as follows:
\begin{equation} \label{eq:more}
\mathbf{g}_{u}^{k-1}(i) = \sigma ( w_u^{k-1}(i, 1) \cdot \mathbf{h}_{u}^{k-1} + \sum_{r=1}^{R-2} w_u^{k-1}(i, r+1) \cdot \mathbf{h}_{\mathcal{N}^{(r)}(u)}^{k-1})
\end{equation}
where $\mathbf{h}_{\mathcal{N}^{(r)}(u)}^{k-1}$ denotes the aggregated vector through the $r$-th type of neighborhood according to $\mathbf{S}_{U}^{(r)}$. Likewise, the weights $\{w_u^{k-1}(i, r+1)\}_{r=0}^{r=R-2}$ of local filter $i$ are shared on the graph. The purpose of these weights is to learn the contribution of different types of neighborhoods. For instance, how does the user-user (by query words) relationship affects the final representation? 

\begin{figure} 
\includegraphics[height=1.8in, width=3.4in]{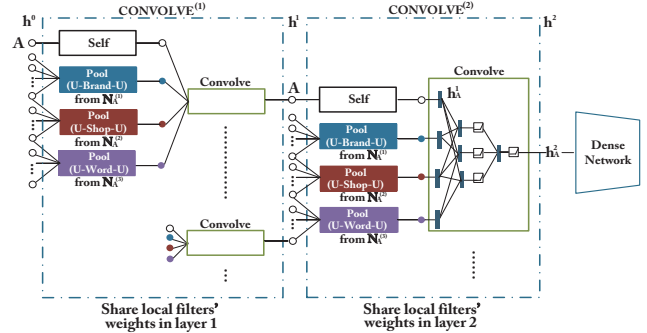}
\caption{Overview of our IntentNet model}
\label{fig:intentnet}
\end{figure}

An intuitive overview of the generalized IntentNet model with two convolutional layers is provided in Fig.~\ref{fig:intentnet}. A ``Convolve'' box performs the convolution function for one node on the graph (Detailed structure of a convolve box is shown in the second layer). It takes the node itself and its pooled heterogeneous neighborhoods as input. In each layer, convolution weights are shared among nodes.


\subsection{Dual Graph Convolution in HIN}
\label{sec:dual}
To handle the heterogeneity between users and items, we propose a dual graph convolution model to learn their embeddings. We use $\mathbf{x}_u$ and $\mathbf{x}_v$ to represent the input feature vectors of user $u$ and item $v$, respectively. In addition, we also sample a negative item for each user-item link to form a complete training tuple as $(\mathbf{x}_u, \mathbf{x}_v, \mathbf{x}_{neg})$, where the negative item is served as a contrast for the positive item in the training process.

We employ two IntentNets, $\textrm{IntentNet}_u$ and $\textrm{IntentNet}_v$, for users and items respectively. By iteratively running $q$ times of convolutional forward propagation as in Eq~(\ref{eq:aggregation}), Eq~(\ref{eq:more}) and Eq~(\ref{eq:kernel}) and additional dense forward propagation via the fully-connected layers, we can obtain the final user and item representations $\mathbf{z}_u$, $\mathbf{z}_v$, by $\textrm{IntentNet}_u$ and $\textrm{IntentNet}_v$ respectively. Although there is semantic gap between user space and item space, the additional three dense layers of IntentNet can play a role in projecting both users and items into a common embedding space. Besides, we also obtain $\mathbf{z}_{neg}$ for the sampled negative item in a training tuple by $\textrm{IntentNet}_v$. 

For training the parameters of the model, we minimize the following triplet loss function:
\begin{equation} \label{eq:loss}
\mathcal{J}(\mathbf{x}_u, \mathbf{x}_v, \mathbf{x}_{neg}) = \max \lbrace 0, \mathbf{z}_u \cdot \mathbf{z}_{neg} - \mathbf{z}_u \cdot \mathbf{z}_{v} + \delta \rbrace
\end{equation}
This triplet loss is designed as a max-margin approach, where $\delta$ denotes the margin hyper-parameter, and inner product is used to measure the similarity score between a user node and an item node. The core idea is that the score between the user and the linked item should be higher than that between the user and a sampled negative item. Minimizing Eq.~(\ref{eq:loss}) can actually maximize the margin of these two scores, which results in a model where high scores can probably lead to real connections.


Moreover, in order to train a robust model to distinguish positive items from negative items that are similar, we sample negative items in the same root category as the corresponding positive items to ensure ``hardness'' of the learning. More details of negative sampling are described in Appendix~\ref{sec:negative-sampling}. 

\subsection{The IntentGC Framework}
\label{sec:framework}

We now summarize our solution in a framework for user-item recommendation, as illustrated in Algorithm \ref{algorithm:IntentGC}.

\makeatletter  
\def\BState{\State\hskip-\ALG@thistlm}  
\makeatother  

\begin{algorithm}[h]
  \caption{IntentGC}
  \label{algorithm:IntentGC} 
  \begin{algorithmic}[1]
    \State \textbf{Network Translation}:
    \State \quad Translate the original HIN $\mathcal{G} = (\mathcal{V}, \mathcal{E})$ into user-item HIN $G$ according to Section~\ref{sec:translation}
    
    \State \textbf{Training}:
    \State \quad Initialize all parameters of IntentNets: $\mathbf{\Omega}_u$ and $\mathbf{\Omega}_v$
    \State \quad Obtain feature matrices $\mathbf{X}_u = [\mathbf{x}_u]$ and $\mathbf{X}_v = [\mathbf{x}_v]$
    \State \quad For $i=1$ to $batch num$:
      \State \quad \quad Sampling mini batch of tuples <$\mathbf{x}_u$, $\mathbf{x}_v$, $\mathbf{x}_{neg}$>
      \State \quad \quad Generate representations for each tuple according to Section~\ref{sec:intentnet}. In real implementation, this is computed via matrix operation:
      \State \quad \quad \quad Set $\mathbf{z}_u = \textrm{IntentNet}(\mathbf{x}_u, \mathbf{\Omega}_u)$
      \State \quad \quad \quad Set $\mathbf{z}_v = \textrm{IntentNet}(\mathbf{x}_v, \mathbf{\Omega}_v)$
      \State \quad \quad \quad Set $\mathbf{z}_{neg} = \textrm{IntentNet}(\mathbf{x}_{neg}, \mathbf{\Omega}_v)$
      \State \quad \quad Update $\mathbf{\Omega}_u$ and $\mathbf{\Omega}_v$ by minimizing the triplet loss in Eq.~(\ref{eq:loss})
        
    \State \textbf{Inference}:
      \State \quad $\mathbf{Z}_u = \textrm{IntentNet}(\mathbf{X}_u, \mathbf{\Omega}_u)$ using trained $\mathbf{\Omega}_u$
      \State \quad $\mathbf{Z}_v = \textrm{IntentNet}(\mathbf{X}_v, \mathbf{\Omega}_v)$ using trained $\mathbf{\Omega}_v$
      \State \quad Obtain $\widehat{\mathcal{E}}^{p}_{label}$ by approximate $K$-nearest search according to $\mathbf{Z}_u$ and $\mathbf{Z}_v$
  \end{algorithmic}  
\end{algorithm}

The IntentGC framework has three main components: 1) Network Translation, 2) Training, 3) Inference. We provide more details in the following:

\vspace{3mm} \noindent \textbf{Network Translation}. The input of our algorithm is a heterogeneous information network $\mathcal{G}$ which is constructed by historical records. Following the method described in Section~\ref{sec:translation}, we generate second-order relationships by auxiliary nodes and translate the origin HIN $\mathcal{G}$ into the user-item HIN $G$ (line 2).

\vspace{3mm} \noindent \textbf{Training}. Given the translated graph $G$, we train our model in four steps: 1) Initialization. We initialize all parameters of the model and obtain feature vectors for users and items (lines 4-5). For clarity, $\mathbf{\Omega}_u$ (or $\mathbf{\Omega}_v)$ denotes the matrix of all parameters of $\textrm{IntentNet}_u$ (or $\textrm{IntentNet}_v$). $\mathbf{X}_u = [\mathbf{x}_u]$ and $\mathbf{X}_v = [\mathbf{x}_v]$ denotes the feature matrices for all users and items, respectively; 2) Sampling. We generate <$\mathbf{x}_u$, $\mathbf{x}_v$, $\mathbf{x}_{neg}$> tuples in a mini-batch manner, where each <$\mathbf{x}_u$, $\mathbf{x}_v$> pair is sampled from user-item edges and $\mathbf{x}_{neg}$ is a negative item sampled in the same root category (line 7); 3) Forward propagation. We feed the mini-batch to $\textrm{IntentNet}_u$ and $\textrm{IntentNet}_v$ and obtain the output representation vectors (lines 8-11). Each IntentNet contains $q$ graph convolutional layers for content propagation and three fully-connected layers for capturing feature interactions; 4) Parameter Updating. We update the parameters of model by performing gradient descent to minimize the triplet loss function in Eq.~(\ref{eq:loss}) (line 12). Steps 2-4 are performed iteratively until stopping condition is met. 

\vspace{3mm} \noindent \textbf{Inference}. After training, we can process all users and items to get their $\mathbf{Z}$ vectors (lines 14-15), and perform approximate $K$-nearest neighbors search \cite{andoni2006near} accordingly for recommendation (line 16). 


\section{Experiments}
To demonstrate the effectiveness of our methods for user-item recommendation, we conduct a comprehensive suite of experiments on two large-scale datasets for offline evaluation (Section~\ref{sec:offline}), and also via online A/B test on the advertising platform of Alibaba (Section~\ref{sec:online}). We aim to answer the following questions: i) How does IntentGC perform compared to state-of-the-art recommendation algorithms? ii) Is the proposed IntentNet model more efficient and effective than GraphSage on billion-scale graph? iii) Can multiple kinds of auxiliary relationships further improve the performance?


\subsection{Datasets}
We use two real-world datasets with timestamps in offline evaluation: 1) a dataset of 9 days' records extracted from Taobao App of Alibaba, denoted as Taobao; 2) a public dataset with product reviews provided by Amazon, denoted as Amazon\footnote{http://jmcauley.ucsd.edu/data/amazon/}. In Taobao dataset, we treat user clicks on items as explicit preferences, so the task is to recommend a user a list of items that he would probably click. In Amazon dataset, we consider high ratings as explicit preferences on items, so the prediction task is to recommend a user a list of items that would probably receive high ratings from him. The statistics of the two datasets are shown in Table~\ref{tab:statistics}. In online evaluation, for each day we will use a dataset with similar format of Taobao, except that it contains the latest records of the last week. We leave the details of data preparation in Appendix~\ref{sec:dataprocessing}.

\begin{center}
\begin{table} \footnotesize
  \caption{Statistics of two datasets}
  \label{tab:statistics}
  \begin{tabular}{|c|c|c|c|c|}
    \hline
    \textbf{dataset} & \textbf{\#users} & \textbf{\#items} & \textbf{\#labeled edges} & \textbf{\#auxiliary edges} \\
    \hline
    Taobao & 278M & 250M & 1.8 billion & 44.1 billion \\
    Amazon & 14M & 6M & 12 M & 37 M \\
  \bottomrule
\end{tabular}
\end{table}
\end{center}

\subsection{Compared Methods}
The compared algorithms in our experiments include:

\vspace{0.5mm} \noindent \textbf{DeepWalk}: This is a classic homogeneous network embedding method. We implement a similar approach like \cite{wang2018billion} for item-item recommendation. 

\vspace{0.5mm} \noindent \textbf{GraphSage}: This is a state-of-the-art method of GCN. Like PinSage in \cite{ying2018graph}, we implement it for item-item recommendation. 

\vspace{0.5mm} \noindent \textbf{DSPR}: This method is a Deep Semantic Similarity Model (DSSM) based algorithm adopted by many companies \cite{xu2016tag}, including Alibaba. It utilizes only explicit preferences in the model.

\vspace{0.5mm} \noindent \textbf{Metapath2vec++}: This is a widely used heterogenous network embedding algorithm \cite{dong2017metapath2vec}. We implement it on the translated graph $G$ and utilize user-user-item-item as the meta-path to obtain representations.

\vspace{0.5mm} \noindent \textbf{BiNE}: This method preserves both explicit preferences and one type of auxiliary relationships by joint optimization (regularization) \cite{gao2018bine}. We implement it with the deduced user-user/item-item relationships as in \cite{gao2018bine}, which can also be viewed as auxiliary relationships.

\vspace{0.5mm} \noindent \textbf{IntentGC(Single)}: This is a simple version of IntentGC that we propose in this paper. In this version, we capture only one type of auxiliary relationships (the same one with BiNE) for comparison.

\vspace{0.5mm} \noindent \textbf{IntentGC(All)}: This is the full version of IntentGC. In this version, we incorporate all kinds of heterogeneous auxiliary relationships.

There are actually three types of methods in our experiments: 1) DeepWalk and GraphSage are item-item recommendation methods which do not capture user information in the model. 2) DSPR is a user-item recommendation method that directly predicts user preferred items but consider no auxiliary relationships. 3) Metapath2vec++, BiNE, and our methods not only take advantage of explicit preferences but also auxiliary relationships. This setting provides us a comprehensive study of our methods as well as insights of progress in recommendation.


\subsection{Offline Evaluation}
\label{sec:offline}
To evaluate the performance, we employ two metrics: 1) area under the ROC curve (AUC) and 2) mean reciprocal rank (MRR), which takes into account the rank of item $i$ among recommended items for user $u$:
\begin{equation} \label{eq:mrr}
MRR = \frac{1}{|\mathcal{L}|} \sum_{(u, i) \in \mathcal{L}} \frac{1}{\lceil R_{u, i} / 100 \rceil}
\end{equation}
where $R_{u, i}$ is the rank of the item $i$ in the recommended item list for user $u$, $\mathcal{L}$ is the set of positive user-item pairs (i.e., clicks in Taobao, high ratings in Amazon) in the test dataset. Due to the large scale of testing instances (over 3 billion in Taobao), we employ a scaled-version of the MRR in Eq.~(\ref{eq:mrr}) by the factor 100 as in \cite{ying2018graph}.

\vspace{3mm} \noindent \textbf{Performance Comparison}. The performance results of all the methods are presented in Table~\ref{tab:performance}. In Taobao dataset, it shows that our proposed IntentGC algorithms outperform state-of-the-art algorithms significantly in both AUC and MRR. There are four important observations in Table~\ref{tab:performance}: 1) from DeepWalk to GraphSage, the AUC increases 0.030 and MRR increases 44.2\%, indicating that the content propagation power of GCN is quite useful for recommendation; 2) from GraphSage to DSPR, the AUC increases 0.0198 and MRR increases 58.6\%. It presents the fact that predicting user-item link directly is much more appropriate than homogeneous network embedding methods in the task of user-item recommendation. 3) from BiNE to IntentGC(Single), the AUC increases 0.015 and MRR increases 41.5\%. This demonstrates that, with the same type of auxiliary relationships (deduced user-user/item-item relationships), IntentGC can learn better than BiNE, which confirms our analysis in the beginning. 4) from IntentGC(Single) to IntentGC(All), the AUC increases 0.012 and MRR increases 37.8\%, which demonstrates that modeling heterogeneous auxiliary relationships can further improve the performance. Moreover, IntentGC(ALL) also outperforms DSPR. Because information of all auxiliary objects are also designed in the input features shared among the methods, IntentGC(ALL) could be viewed as reducing to DSPR when no auxiliary relationships are used. This suggests that further employing the information of auxiliary objects as heterogeneous auxiliary relationships in IntentGC can boost the performance. From these observations, we can gain some insights of progress in the recommendation literature. Similar results are also observed in the public Amazon dataset in Table~\ref{tab:performance}.


\begin{table} \footnotesize
  \caption{Offline performance of compared methods}
  \label{tab:performance}
  \begin{tabular}{ccccc} 
    \toprule
    dataset & Taobao & Taobao & Amazon & Amazon \\
    algorithm & AUC & MRR & AUC & MRR \\
    \midrule
     DeepWalk & 0.622829 & 0.0822 & 0.675525 & 0.6230  \\
     GraphSage & 0.653121 & 0.1186 & 0.716853 & 0.7847 \\
    \midrule
     DSPR & 0.672956 & 0.1881 & 0.778336 & 1.2102 \\
    \midrule
     metapath2vec++ & 0.673261 & 0.1893 & 0.783334 & 1.3325 \\
     BiNE & 0.674835 & 0.1920 & 0.789051 & 1.4693 \\
    \midrule
     \textbf{IntentGC(Single)} & \textbf{0.689367} & \textbf{0.2718} & \textbf{0.826094} & \textbf{2.2249} \\
     \textbf{IntentGC(All)} & \textbf{0.701740} & \textbf{0.3746} & \textbf{0.837589} & \textbf{2.7981} \\
  \bottomrule
\end{tabular}
\end{table}


\vspace{3mm} \noindent \textbf{IntentNet vs. GCN}. In order to scale up the graph convolution for billion-scale user-item graphs, we propose IntentNet with the idea of dividing the functionality of GCN into two separate components (vector-wise convolutional network on graphs and dense network on nodes). In this way, time complexity can be reduced dramatically. Moreover, by avoiding unnecessary interactions in the model, less parameters in IntentNet would better resist overfitting. To test these ideas, we evaluate two IntentGC variants to model only one type of auxiliary relationships in the experiment: one with IntentNet as the model and the other with GraphSage (a state-of-the-art GCN model) as the model. We denote them as ``IntentGC(Single) with IntentNet'' and ``IntentGC(Single) with GraphSage'', respectively. Based on the same number of training epochs, it can be seen from Table~\ref{tab:intentnet} that IntentGC(Single/All) with IntentNet takes 19/21 hours while IntentGC(Single) with GraphSage takes 78 hours. Thus, IntentGC(Single/All) with IntentNet can accomplish training in one day with a dataset of records from the past 7 days (i.e., the size of the training set in Taobao dataset) in order to serve online for tomorrow. However, IntentGC(Single) with GraphSage trained on day $T$ can only be deployed online on day $T+5$, since it needs more than 3 days of training. Because user-item recommendation is highly sensible to time, IntentGC(Single) with GraphSage is not feasible in practice. In our work, IntentNet also enables us to handle more heterogeneous relationships with a similar training time.

Although IntentGC(Single) with GraphSage cannot provide a trained model on $T+1$ in reality, we still use the dataset of day $T+1$ to evaluate its performance for fair comparison. In Tables~\ref{tab:intentnet} and~\ref{tab:intentnet-amazon}, we observe that IntentGC(Single) with IntentNet is also more effective than IntentGC(Single) with GraphSage in both Taobao and Amazon. It suggests that vector-wise convolution on graphs actually avoids useless feature interactions and fits the data better without overfitting.

\begin{table} \footnotesize
  \caption{IntentNet vs GCN in Taobao}
  \label{tab:intentnet}
  \begin{tabular}{cccc} 
    \toprule
    algorithm & Training Time & AUC & MRR \\
    \midrule
     IntentGC(Single) with GraphSage & 78h & 0.680296 & 0.2226 \\
     IntentGC(Single) with IntentNet & 19h & 0.689367 & 0.2718 \\
     IntentGC(All) with IntentNet & 21h & 0.701740 & 0.3746 \\
    \bottomrule
\end{tabular}
\end{table}

\begin{table} \footnotesize
  \caption{IntentNet vs GCN in Amazon}
  \label{tab:intentnet-amazon}
  \begin{tabular}{cccc} 
    \toprule
    algorithm & AUC & MRR \\
    \midrule
     IntentGC(Single) with GraphSage & 0.808307 & 1.7212 \\
     IntentGC(Single) with IntentNet & 0.826094 & 2.2249 \\
    \bottomrule
\end{tabular}
\end{table}

\begin{center}
\begin{figure}
\subfigure[Taobao]
{ 
\begin{minipage}[t]{0.45\linewidth}
\includegraphics[height=20mm, width=40mm]{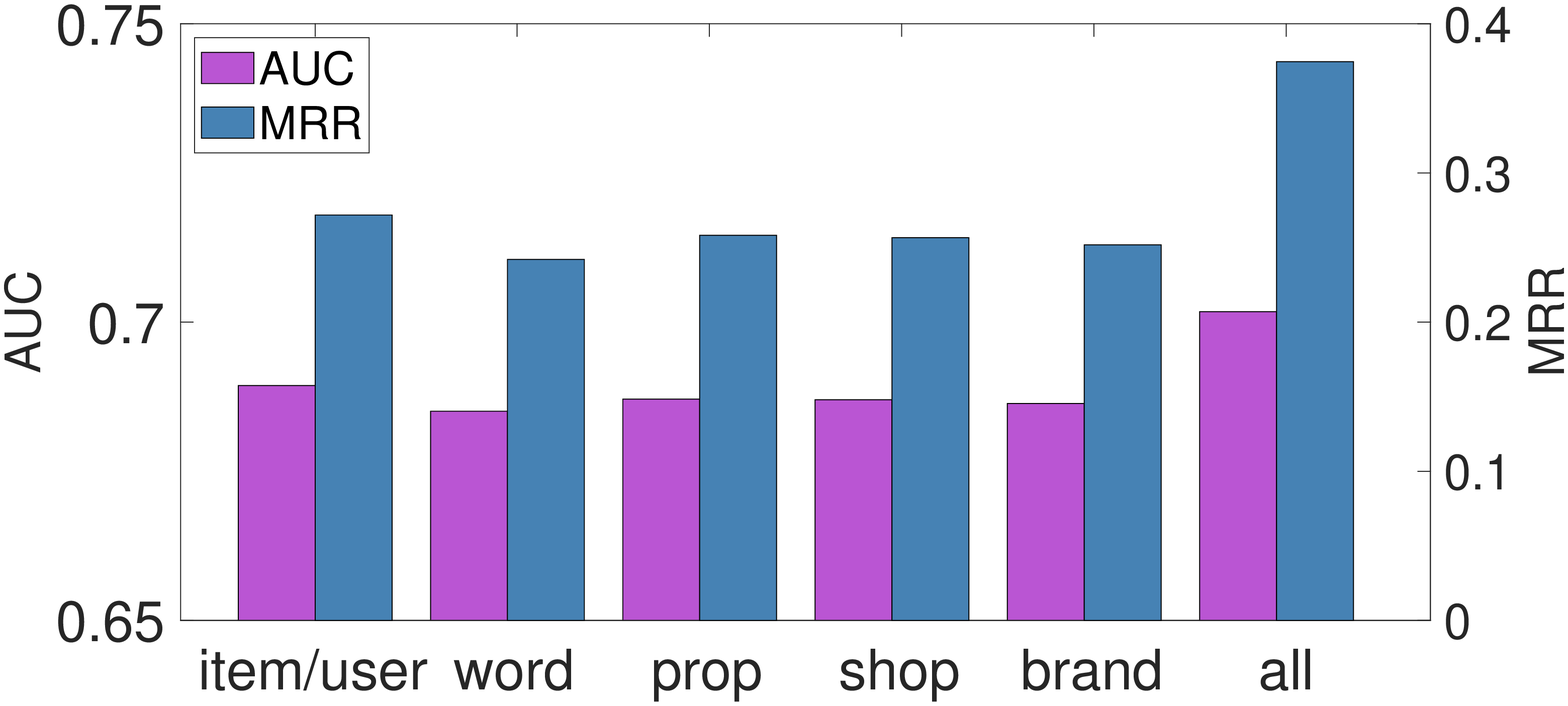}
\end{minipage}
}
\subfigure[Amazon]
{ 
\begin{minipage}[t]{0.45 \linewidth}
\includegraphics[height=20mm, width=40mm]{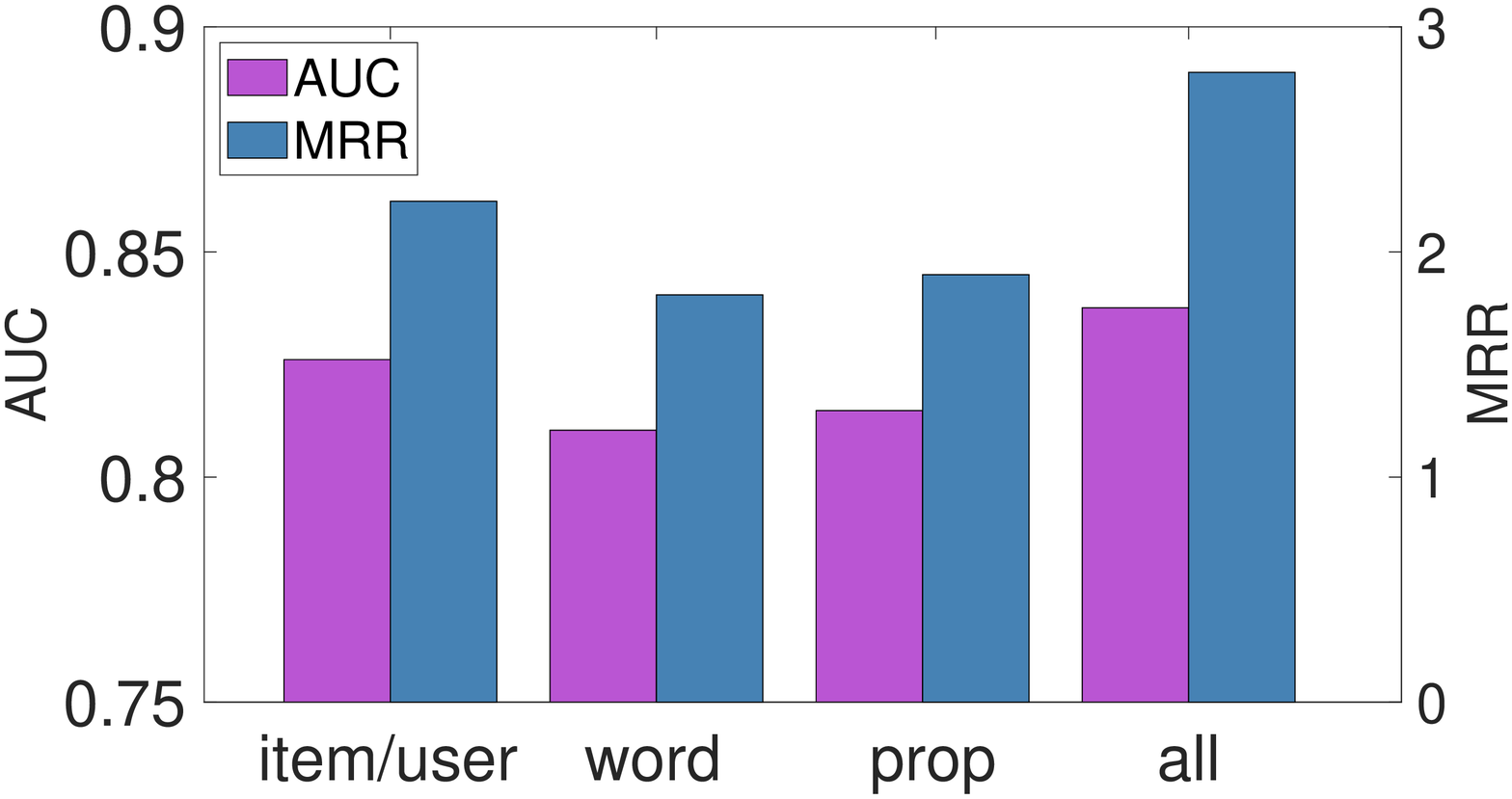}
\end{minipage}
}
\caption{The study of heterogeneous relationships}
\label{fig:study}
\end{figure}
\end{center}

\vspace{3mm} \noindent \textbf{Effectiveness of Different Types of Auxiliary Relationships}. By now, we have only compared one type of auxiliary relationships in IntentGC(Single) with IntentGC(All). To study whether modeling heterogeneous auxiliary relationships can really lead to better performance than any single type, we deploy a specific version of IntentGC(Single) for each type of auxiliary relationships. In Fig.~\ref{fig:study}, we name these specific versions by the type name of the objects used for generating auxiliary relationships. For example, ``word'' in Fig.~\ref{fig:study} means we only use the auxiliary relationships generated by user-word-user and item-word-item links, ``item/user'' in Fig.~\ref{fig:study} means we use the same auxiliary relationships with BiNE. It shows that, the IntentGC(All) outperforms all single type versions in both Taobao and Amazon, which means that heterogeneous auxiliary relationships can complement each other and are able to improve performance when modeled together.

\subsection{Online Evaluation}
\label{sec:online}
Based on the same algorithms we deploy a recommender system on Alibaba Advertising Platform. In advertising, we recommend ads (items) to users. This section presents the results of online A/B tests in the Alibaba Advertising Platform. To evaluate different algorithms, we use click-through rate (CTR) as the performance metric, which is the key objective for advertising. A higher CTR means a better advertising. It should be noted that, due to Alibaba's business policy, we temporarily cannot expose the absolute CTR values. Instead we use a scaled CTR which multiplies the real CTR with a constant coefficient. This will not affect performance comparison.

The results are reported in Table~\ref{tab:online_performance}. We compare the performance of our algorithms to the best baseline that considers only user-item links (DSPR) and the best baseline that leverages both explicit preferences and auxiliary relationships (BiNE). We find that IntentGC(Single) and IntentGC(All) consistently outperform these baseline methods. It can be observed that the IntentGC(Single) improves CTR by 30.4\% compared to the best baseline BiNE, and IntentGC(All) further improves CTR by 26.7\% compared to IntentGC(Single).

\begin{table} \footnotesize
  \caption{Online performance of compared methods.}
  \label{tab:online_performance}
  \begin{tabular}{p{1.7cm}p{1.1cm}p{4.5cm}}
    \toprule
    algorithm & scaled CTR & model capability \\
    \midrule
     DSPR & 3.2812 & consider no auxiliary relationships \\
     BiNE & 3.3822 & consider one type of auxiliary relationships \\
     \textbf{IntentGC(Single)} & \textbf{4.4136} & consider one type of auxiliary relationships \\
     \textbf{IntentGC(All)} & \textbf{5.5964} & consider heterogeneous auxiliary relationships \\
  \bottomrule
\end{tabular}
\end{table}

\section{Conclusions}
In this paper, we propose the first framework to capture heterogeneous auxiliary relationships in recommendation. Empirical experiments demonstrate that these heterogeneous relationships are practically useful in predicting users' preferences and complementary to each other. In our framework, we design a faster graph convolutional network for billion-scale application. Experimental results indicate that our method can avoid unnecessary feature interactions with better effectiveness and efficiency.

In future work, it is worth extending our approaches and ideas to other tasks, such as information retrieval. Besides, it is also meaningful to study a dynamic graph convolutional model to emphasize real-time user behaviors as well as interest drift.

%

%
\bibliographystyle{ACM-Reference-Format}
\balance
\bibliography{sample-base}

%

\newpage

\appendix

\section{Reproducibility}

\subsection{Hyper-parameter Settings}
For reproducibility, we provide the settings of some key hyper-parameter settings in Table~\ref{tab:hyper}. We also have a few insights for hyper-parameter tuning in the following. 1) Compared algorithms are not quite sensitive to learning rate between 0.001 to 0.0001; 2) 100-1000 is a good range for the mini-batch size. In our experiments, we use 200; 3) The standard deviation for initializing parameters is quite important. For network parameters, smaller standard deviation might lead to bad local optimum. For feature embedding parameters, larger standard deviation causes slow convergence; 4) 10 neighbors for convolution are good enough for significantly improving the performance, more neighbors raise the time complexity and the improvement is not so much. For practical usage, we use 10 neighbors in this paper and also in product.

It is worth to note that, we use the same hyper-parameter settings in experiments for all the algorithms, which results in good convergence. This also ensures a fair condition for performance comparison.

\subsection{Hardware and Software}
In distributed learning, we use a cluster of 200 machines for each epoch of each algorithm, where an instance machine has 32 CPU cores and 128G memory. All models are implemented on TensorFlow v1.7 and Python v2.7. The wiki for distributed environment setting is shared in \href{https://github.com/alibaba/euler/wiki}{https://github.com/alibaba/euler/wiki}. With the distributed environment been set, readers can reproduce our results by running codes in \href{https://github.com/peter14121/intentgc-models}{https://github.com/peter14121/intentgc-models}.

\subsection{Dataset Preprocessing}
\label{sec:dataprocessing}

\textbf{Taobao}: We collect all records from October 11, 2018 to October 19, 2018 in Taobao App. It is a dataset with rich information including user profiles, item descriptions, user click behaviors, purchase data, impression lists of pages, etc. In this dataset, we consider user clicks as explicit preferences on items (labeled edges). Using other types of behaviors to be the preference labels is also valid. However, we choose clicks since this kind of information is relatively abundant and is a superset of purchases. To simulate a real recommendation task in offline evaluation, we use the data from October 11 to October 17 as the training set, and use the data in October 19 as the test set (we skip the data for 18th since in practice the model training needs one day). In this way, the offline recommendation is quite similar to the real usage in online product, and share the same evaluation setting with the online A/B tests. 

\vspace{2mm}  \noindent  \textbf{Amazon}: This dataset is a subset of product reviews and metadata provided by Amazon. Since user clicks are not provided in the dataset, we consider high ratings (rating = 5) as explicit preferences on items. With this definition, we can formulate a recommendation task. Because Amazon only opened limited data in the public dataset, we use a wide range of time for both training and test. In the experiment, we use the data from August 1, 2006 to July 30, 2013 as the training set, and use the data from August 1, 2013 to July 23, 2014 as the test set. This still keeps the nature of predicting unseen behaviors in the recommendation task and split ratio is also similar to that of the Taobao dataset. 


\begin{table}[h] \footnotesize
  \caption{Hyper-parameter settings}
  \label{tab:hyper}
  \begin{tabular}{cc} 
    \toprule
    hyper parameter & setting \\
    \midrule
     learning rate & 0.0001 \\
     optimizer & momentum \\
     mini-batch size & 200 \\
     full-connected network structure & [110, 800, 300, 100] \\
     q-stack number & 2 \\ 
     stddev for initializing network & 0.8 \\
     stddev for initializing embedding & 0.4 \\
     $\delta$ in Eq.~(\ref{eq:loss}) & 0.3 \\
     neighbors for convolution & 10 \\
    \bottomrule
\end{tabular}
\end{table}

\subsection{Feature Design}
Since there are some differences between Taobao dataset and Amazon dataset, we design features for each of them. The corresponding features are illustrated in the following:

\vspace{3mm} \noindent \textbf{Features of Taobao}: There are relatively richer information in Taobao dataset since we own the complete records on the platform. Main features for user nodes include: preferred categories (including various levels), favorite brands, frequent query words, user profile (age, occupation, income, province, gender, etc.), member information (vip level, star level, etc.), purchased items' statistics (average price, trade number, etc.), visited shops, preferred properties, etc. Main features for item nodes include: words of title, category path, properties, brand, shop information, statistics (price, trade, CTR, etc.), profiles of frequently visiting users, business types (e.g., b2c or c2c), etc. These are the major features designed for Taobao dataset that can lead to good performance, we omit the other features due to business secret.

\vspace{3mm} \noindent \textbf{Features of Amazon}: We have limited access to Amazon's data, so we designed a smaller feature set for it. Main features for user nodes include: preferred categories (including two levels), favorite brands, preferred words collected from high rated items, high rated items' statistics from metadata, average rating score, consumption level, number of reviews written. Main features for item nodes include: words of title, category path, brand, statistics from metadata (price, reviews, etc.), popular review users' statistics (rating distribution, number of reviews written, etc.). Although we cannot access many other features in Amazon, the major features are similar to the ones of Taobao dataset. For a complete reveal of features, please refer to the code.

We also obtain a helpful insight for feature designing by experiments: Since we model recommendation as a link prediction task on the user-item graph, it is expected that a user and his preferred items should be mapped closer. This inspires us to design co-features for user nodes and item nodes. To be specific, if we use price as a feature for item nodes, it is useful to design a matching feature (e.g., average price) for user nodes as well. This is an effective strategy to design features for user-item recommendation.

For either Taobao and Amazon dataset, there are two types of features: continuous features and discrete features. For continuous features (e.g., CTR), it is straightforward to add them directly to the feature vector. For discrete features, we first construct an ID dictionary for mapping discrete values, and then use the embedding lookup function to obtain the embedding vectors, which will be concatenated into the feature vector. This preprocessing operation is equivalent to feeding a one-hot vector through a fully-connected neural layer for each discrete feature, but can be much faster. If a feature contains multiple discrete values, we use the tf.nn.embedding\_lookup\_sparse function for preprocessing.

\subsection{Negative Sampling}
\label{sec:negative-sampling}
For each labeled edge ($u, v$) that represents $u$'s explicit preference on item $v$, we need to sample multiple negative instances for training, so that we can learn representations by maximizing the relevance between users and positive items while minimizing the relevance between users and negative items. To make the model capable of distinguishing hard cases, we only consider negative items in the same leaf category as the positive item. The steps of negative sampling include: 1) First, we calculate the weights (click frequency in Taobao; review counts in Amazon) of items in each leaf category, so the weights of the same item in different category are also different. 2) Then, for any labeled edge ($u, v$), we pick a negative item in the same leaf category by weighted random selection. For instance, if a category has three items with weights 70, 20, 10, they will be picked with probabilities 0.7, 0.2, 0.1, respectively. 3) Last, if the sampled item together with the target user form a positive instance in the training set, we discard it and do re-sampling. This negative sampling process is run 5 times for each labeled edge. Usually, more negative instances lead to more robust performance but consume more computing resources.

\subsection{Heterogeneous Auxiliary Relationships}
As mentioned in Section~\ref{sec:translation}, robust heterogeneous relationships and neighborhoods are generated according to the second order proximity via auxiliary nodes. However, in large-scale graph, the computation of second order proximity is not feasible. For instance, a hot brand typically has millions of fans, which results in $\mathcal{O}(10^{12})$ user pair counting for that brand. For this reason, we use another practical implementation, in which we do not need to calculate the complete second-order proximity. The basic idea is that, hot brands are actually not useful to capture user relationship, because almost everyone likes them. Hence, if two users prefer a same hot brand, this only provides a very weak evidence that they share similar tastes. Following this idea, we only calculate second-order proximity by normal brands with less than $20$k user visits per day. We find that they are more useful to measure the degree of similarity. For the other auxiliary nodes and edges, we generate second order proximity relationships among users/items in a similar way. We provide the kinds of captured auxiliary relationships in Table \ref{tab:implicit}. It should be noted that, since Amazon only provides a limited access to its data, we can only use a subset of the whole set of auxiliary relationships for Amazon. Nevertheless, all the auxiliary relationships in Table \ref{tab:implicit} can be obtained in the Taobao dataset. In addition, a very few nodes have less than 10 neighbors (or no neighbor). For those nodes, default nodes are generated by replicating its most similar neighbor (or the node itself if no neighbor exists) and used to fill the blanks.

\subsection{System Architecture}
\begin{figure}
\includegraphics[height=2.4in, width=3.4in]{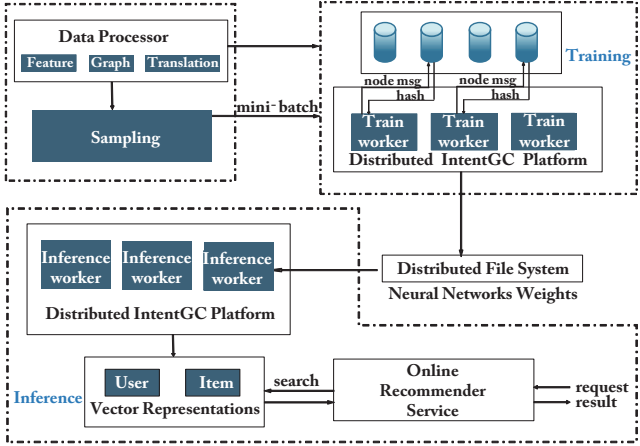}
\caption{System Architecture}
\label{system}
\end{figure}

The system architecture of our approach is depicted in Fig.~\ref{system}, which is implemented to meet the requirements of the IntentGC framework in Section~\ref{sec:framework}. The data preprocessing of features and graphs before model training is handled offline by MapReduce downstream in Java, and both training and inference components are implemented on a distributed IntentGC platform in Python. It is worth to note that, this is a highly flexible implementation in that we remove the limitation of training on clustered mini-graph batches. Instead of producing clustered mini-graphs for every batch, we sample random nodes and fetch their neighborhoods from the graph indexing engine by hash keys in the run time of training. The inference component is much like the training component except without backward propagation. After inference, user representations and item representations are stored in database for online services. All experiments in this paper are implemented on a graph learning framework named Euler. Its source code can be found in \href{https://github.com/alibaba/euler}{https://github.com/alibaba/euler}.


\begin{center}
\begin{table} \footnotesize
  \caption{Kinds of heterogeneous auxiliary relationships}
  \label{tab:implicit}
  \begin{tabular}{|c|c|c|}
    \hline
    \textbf{Type of Schema} & \textbf{Information} & \textbf{Example} \\
    \hline
    user-word-user & semantic desire & office lady style, pet \\
    user-brand-user & brand taste and loyalty & LOUIS VUITTON, APPLE \\
    user-shop-user & browsing habit & specific visited shop list \\
    user-item-user & common interests on item & specific clicked item list \\
    user-property-user & preferred property & cheap, quality, fashion \\
    item-word-item & text similarity & sweet, beautiful \\
    item-property-item & property similarity &  quality, fashion\\
    item-brand-item & brand similarity & LOUIS VUITTON, APPLE \\
    item-user-item & common users & specific coming user list \\
    item-shop-item & belong to the same shop & specific shop \\
  \bottomrule
\end{tabular}
\end{table}
\end{center}

\end{document}